\documentclass[graybox,vecphys]{svmult}

\usepackage{helvet}         % selects Helvetica as sans-serif font
\usepackage{courier}        % selects Courier as typewriter font
\usepackage{type1cm}        % activate if the above 3 fonts are
\usepackage{makeidx}         % allows index generation
\usepackage{graphicx}        % standard LaTeX graphics tool
\usepackage{multicol}        % used for the two-column index
\usepackage[bottom]{footmisc}% places footnotes at page bottom

\makeindex             % used for the subject index
\begin{document}

\title*{Progress in Experimental Measurements of the Surface-Surface Casimir Force: Electrostatic Calibrations and Limitations to Accuracy}
\titlerunning{Surface-Surface Casimir Force Experiments}% for an abbreviated version of
\author{Steve K. Lamoreaux}
\institute{Yale University, Physics Department \email{steve.lamoreaux@yale.edu}}
%\and Name of Second Author \at Name, Address of Institute \email{name@email.address}}
\maketitle

\abstract*{A brief review of recent progress in Casimir force measurements is presented. Emphasis will be placed on recently discovered problems associated with background electrostatic effects.}

\abstract{Several new experiments have extended studies of the Casimir force into new and interesting regimes. This recent work will be briefly reviewed.  With this recent progress, new issues with background electrostatic effects have been uncovered.  The myriad of problems associated with both patch potentials and electrostatic calibrations are discussed and the remaining open questions are brought forward.}

\section{Introduction}
\label{sec:1}
Nowadays, it is unclear what it means to write a review article, or a review chapter for a book, on a particular subject.  This unclarity results simply from the ease with which modern digital reference and citation resources can be used; with a mere typing of a keyword or two into a computer hooked up to the internet, one has an instant review any field of interest.  As such, at the present time, review articles tend to be op-ed pieces that tend to be less than scientifically enlightening.  Rather than continue in the tradition of collecting up a series of electronic database searches, I will give an overview of some recent experiments and also describe how anomalous electrostatic effects might have affected the results of these experiments. This Chapter is not meant to be a review of every paper in the Casimir force experimental measurement field, but a review of what I consider are the credible experiments, that have carried the field forward, that were performed over the last decade or so.  As such, there will be little mention of experimental studies that have claimed 1\% or better agreement, simply because it is unclear to me what these experiments really mean.  If the reader is interested, a recent review of this 1\% level work is presented in \cite{mohideenrev}. Of course I admit freely that my review presented here reflects my own opinions, however I hope the reader accepts or rejects my points based on verifiable facts and an independent scientific analysis.  It must be remembered that simply because a paper appears in print, in a credible and leading journal, it is not necessarily scientifically correct or accepted by the community at large.  Neither does the fact that work is funded by the DOE, NSF, or DARPA (or other funding agencies beyond the realm of the U.S.A.) guarantee its validity or broad acceptance in the scientific community.  An perhaps most interestingly as a remark on the general history of science, the ``consensus opinion" is not necessarily correct either.  In particular, in the surface-surface Casimir force measurement field, there have been more than a few ``Comments" on various papers; the interested reader would do well to ignore most, but not all, of these ``Comments" as they are confusing, if not bogus, but certainly inflammatory.

Watching the field develop since my 1997 experimental result \cite{skl97prl}, which served as a watershed for new interest in surface-surface Casimir force measurements, has been fascinating.  I had no preconceived notions as to how large or small the effect should be relative to the case of assumed simple perfect conductors (e.g., ignoring effects like surface plasmons), but I had no illusions as to the accuracy of my work, hence the words ``Demonstration of the Casimir force" in the title of my paper.  I simply did not have the time or resources to perform a study of possible systematic effects that likely limited the accuracy of my result; the precision was at the 5\% level, at the point of closest approach. Again the accuracy of my result was, and remains, an open question, as it does for any experiment.

At the time the work reported in \cite{skl97prl} was performed, there were no precision calculations of the Casimir force for real materials.  Describing the metal plates with the simplest plasma model, for parallel plates, the correction to the force compared to the perfect conducting case is \cite{schwinger}
\begin{equation}\label{plascorrect}
\eta(d)=1-{16 \over 3}{c\over \omega_p d} ,
\end{equation}
where $\eta(d)$ is a force correction factor which varies with plate separation $d$, $c$ is the velocity of light, and $\omega_p$ is the plasma frequency, where the form of the permittivity of the metal is
\begin{equation}\label{plas}
\epsilon(\omega)=1-{\omega_p^2\over \omega ^2} ,
\end{equation}
which is valid at high frequency. As $\omega$ approaches zero, Eq. (\ref{plas}) become invalid, and in addition the effect of static conductivity must be included also. Equation (\ref{plascorrect}) can be easily modified for a sphere-plane geometry \cite{skl97prl}.  However, the magnitude of this correction
was certainly outside what was reasonable based on the precision of my experiment, which appeared to be best described by plates with perfect conductivity.  There was some skepticism regarding the lack of a finite conductivity correction in my result, and although several theorists expressed interest in performing a more accurate calculation, none did.  Eventually I attempted the calculations myself, with mixed results. My calculations were based on published optical properties of Au and Cu, with the Cu calculations intended as a test case.  These calculations showed  roughly 10-15\% (for Cu) and 20-30\% (for Au) reductions in force, compared to perfect conductors, for distances of order one micron; I eventually found an error in the radius of curvature of the spherical-surface plate used in my experiment \cite{sklcalc, sklerratum} that lowered the experimentally measured force by 10\%, but did not bring the experimental result into agreement with my Au calculation.  Later work showed that Au and Cu are nearly identical, with my Cu result being the more accurate; the discrepancy was due to the way I interpolated between data points in the tabulated optical data \cite{lambrecht1}.  With the refined calculation, my experiment and theory appeared to be in agreement, however by this time I was skeptical of my results, as stated in the ensuing discussion, in \cite{lambrecht2}. Interestingly enough, I had spent considerable effort trying to find corrections that would bring my experimental result into agreement with my original inaccurate calculation, so I felt that I was prepared to comment  against a new theoretical result, obtained by Bostr\"om and Sernelius \cite{bossern}, that leads to a major correction to the Casimir force between real, non-superconducting materials.  This correction reduces the force by a full factor of two at large separations.  More will be said of this correction later in this review; in particular, in light of new electrostatic systematic effects that have recently been discovered, the rhetoric against the result of Bostr\"om and Sernelius no longer appears as certain.  In addition, all of the 1\% work that was reported before \cite{bossern} does not show the predicted correction,  nor does subsequent 1\% level work. So we are faced with the possibility that the degree of precision isn't as high as stated in the 1\% work, or that the theory is not at all understood.  Instead of questioning experimental accuracy, new fantastic theoretical suggestions have been made, regarding the low frequency permittivity of metals, that eliminate the new correction.  This remains a major open topic in the field.

There is a tendency among workers in this field to confuse precision with accuracy, of which I am guilty myself.  {\em Precision} relates to the number of significant figures a measurement device or system provides; lots of digits can be useful for detecting small changes in some ``large" parameter, assuming that the system is stable.  {\em Accuracy} is the assignment of meaning to precision, it is the connection between accepted definitions of, for example, lengths, voltages, and forces, and the measurements that come out of an experimental apparatus.  As an example, for Casimir force measurements using the sphere-plane geometry, an essential parameter is the radius of curvature of the sphere.  A radius of curvature accuracy of 0.5\% for a sphere of 0.2 mm diameter corresponds to 1000 nm, a bit larger than the wavelength of visible light.  Thus optical measurements of adequate accuracy appear as hopeless; can electron microscopy attain this level of precision?  The answer is not obvious. Of course, an experiment can be designed that does not require a high accuracy radius of curvature measurement, {\it e.g.}, when the ratio of Casimir to electrostatic force is measured. Nonetheless, the attention to this problem in those works reporting 1\% or better accuracy does not appear as sufficient to warrant such accuracy claims.  The precision might be that level, but the cross checks required for accurate work are missing.

In general, to attain a given experimental accuracy, say 1\%, requires that the calibrations and force measurements must be done to much better than 1\% accuracy, particularly for comparisons between theory and experiment with no adjustable parameters. As there are possibly five or more absolute measurements that must be made to interpret an experiment, a reasonable requirement for the average calibration accuracy is 0.5\%, assuming that the uncertainties can be added in quadrature (this point is open to debate; many precision measurement experts insist that the uncertainties be simply added, which bring the required average accuracy to the 0.2\% level).  Some of the required calibrations are as follows: The optical properties of the surfaces must be adequately characterized to allow calculation of the force to 0.5\% accuracy;  The radius of curvature of the spherical surface (for a sphere-plane experiment) needs to be measured to 0.5\% accuracy;  The  absolute separation must be determined to high accuracy.  This last point is perhaps the most difficult, as
\begin{equation}
\left|{\delta F\over F}\right|=\left|n{\delta d\over d}\right| ,
\end{equation}
where $n$ is the exponent in the power law.  For a sphere-plane geometry where $n \approx -3$ we see immediately that if we want 0.5\% force accuracy as limited by the distance measurement,  at the point of closest approach, say 100 nm, then the fractional error must be $0.5\%/3$ or about 0.17\%, and when $d=100$ nm this corresponds to $\delta d= 0.17$ nm $=1.7$\space \AA.  This is at the level where, in the atomic force microscopy (AFM) community, the definition of the surface location is agreed as controversial. So we see immediately that it pointless to include any discussion of experiments that claim 1\% accuracy as the radius measurement is not discussed in sufficient detail in any of the papers making such claims.  My statements here should be considered as a call for details.

The general experimental techniques used in all Casimir experiments to date are rather straightforward.  Many experiments employ AFM or micromechanical techniques drawn from fields that enjoy tremendous engineering support.  The trick of Casimir force measurements lies in the attainment of very high force measurement sensitivity subjected to precise and rigorous calibrations, and in the elimination of long-range background electrostatic effects that can mask or distort the now-well-studied AFM signals extrapolated to very large distances.   At large distances, the attractive force between two surfaces, ``the" Casimir force, becomes a property of the bulk material(s) that the plates comprise, and is viewed as a fundamental physical effect arising from the quantum vacuum, as opposed to AFM signals used to detect surface roughness, for example.  Experimental rigor is required to transform precision into accuracy on the fundamental vacuum effect.

Because the measurement techniques are largely borrowed from other fields, I will not give a nuts and bolts discussion of measurements in this review, for the simple reason that I know nothing about AFM techniques. Nowadays one can simply buy an AFM system from Veeco, for example, and adapt it to the samples and longer distance ranges required for Casimir measurement.  There are companies that commercially produce bare cantilevers, and most engineering schools have fabrication facilities where NEMs and MEMs systems can be produced with just about any desired properties in configurations limited only by the imagination.  Alternatively, my own work employs torsion balances, and the interested reader can refer to Cavendish's experiment for most details of such systems.  An analysis of the force sensitivity of a torsion pendulum can be found   in \cite{buttlam}.

The principle advantage to AFM type or torsion pendulum type measurements (in fact there is no fundamental difference between them, it's a matter of scale) is elimination of stiction associated with the fulcrum type balances used in practically all earlier experiments.  The proliferation of high accuracy mechanical and optomechanical translation stages, together with high quality digital data acquisition systems has made precision Casimir force measurement possible; the questions of accuracy are now the central theme, not the simple detection of the force.

This is not to say that the experiments are easy or simple; again, the art of the experiments lies in the attainment of high force measurement sensitivity, reliable calibrations, the production of well-characterized optical surfaces, and the elimination of background effects due to, for example, electrostatic effects.  The electrostatic effects are common to all experiments, either in regard to system calibrations or systematic background effect, or both.  Given the importance of electrostatic effects, I will discuss them at length in this review.

It is often said that the Casimir force is simply the retarded van der Waals potential.  This view strikes me as fundamentally flawed, as the Casimir force does not depend on the properties of the individual atoms of the plates, but on their bulk properties.  Indeed, the non-additivity of the van der Waals effect has been discussed at length in the literature (see \cite{milonni} for a discussion and references).  It is more profitable to think of the Casimir force as the zero point electromagnetic field stress on a parallel plate waveguide.  This force is apparently largest when the waveguide is constructed from perfectly conducting material(s).  The effects of imperfect conductivity can be calculated provided the optical constants of the material(s) are known over an adequate wavelength range. Furthermore, most of the surface-surface Casimir effect is due to conduction electrons.  It is meaningless to assign a retarded van der Waals force between the individual electrons in a conductor.  Likewise, if the Casimir force was simply the retarded van der Waals force, it would make little sense consider  modifying the Casimir force, in a fundamental way, by altering the mode structure imposed by specially tailored boundaries.

\section{Motivation for the Experimental Study of the Casimir Force: Some Recent Results}
\label{sec:2}
The Casimir force is of fundamental interest in that it is taken as evidence for the existence of the fluctuations associated with the quantum vacuum \cite{cas}.  One can almost as easily derive the Casimir force by treating the electromagnetic field classically, with the field fluctuation due to dissipation in the material bodies; this is the Lifshitz approach \cite{lifshitz}.  A principal controversy associated with the quantum vacuum interpretation lies in the fact that the zero point electromagnetic field energy, when integrated to the Planck scale (which is the natural cutoff), leads to a cosmological energy density some 130 orders of magnitude larger than observed. This is an open problem in modern physics.

There are three principal motivations for studying the Casimir force.  One question is how well do we understand the basic underlying physics?  This relates to second motivation which lies in the testing for the existence of short range corrections to gravity, or a new force associated with axion exchange, for example.  For such tests, the Casimir force represents a systematic background effect that must be characterized or physically eliminated by employing a shield.  The third motivation comes from interest in modifying the Casimir force to eliminate stiction, for example, or make it useful in nanodevices.   These categories are not mutually exclusive, and of course overlap considerably as the questions all have a fundamental element.

\subsection{Progress in Understanding the Fundamental Casimir Force}
\label{subsec:1}

In 2000, Bost\"om and Sernelius \cite{bossern} put forward the first fundamentally new idea relating to the surface-surface Casimir effect in over 40 years, since Lifshitz's paper \cite{lifshitz}, which lies in the treatment of  material permittivities in the zero-frequency limit.  The problem of finite conductivity was addressed earlier by Hargreaves and later by Schwinger et al. \cite{schwinger} who proposed a possible means to deal with it, that is, to let the surface material permittivity diverge before setting the frequency to zero.  The point is that in calculating the Casimir force at finite temperature, the integral includes a Boltzmann's factor which accounts for the thermal population of the electromagnetic modes,
\begin{equation}\label{poles}
N(\omega)+{1\over 2}= {1\over e^{\hbar \omega/k_b T}-1} +{1\over 2}= {1\over 2}\coth{\hbar\omega\over 2k_b T} ,
\end{equation}
where $\hbar\omega$ is the energy of a photon, $k_b$ is Boltzmann's constant, and $T$ is the absolute temperature.  Because $\coth x$ has simple poles at $x=\pm i n\pi$, the integral over frequency in calculating the Casimir force can be replaced by a sum of the residues at the poles of Eq. (\ref{poles}), or Matsubara frequencies,
\begin{equation}
\omega_n=  {n\pi k_b T\over \hbar}.
\end{equation}
Analytic continuation of the permittivity function allows the transformation of the integral from over real frequencies to a contour integral on the complex frequency plane, and it is valid to replace the integral over frequency with a sum over the poles.   The upshot is that the transverse electric ($TE$) mode with $n=0$ does not contribute to the force at all if the permittivity diverges slower than $\omega^{-2}$ in the limit as $\omega$ goes to zero.  It is generally assumed that for metals with a finite conductivity,  at zero frequency the permittivity goes as
\begin{equation}\label{cond}
\epsilon(\omega)={4\pi i\sigma\over  c\omega} ,
\end{equation}
in which case the $TE$ $n=0$ mode does not contribute at all to the force.  This is important because at room temperature, at distances greater than about 10 microns, this mode accounts for roughly half of the force.  The implied correction at separations of 1 micron is about 30\%.  This appears to be at odds with a number of experiments, including my own.  In particular, I had spent much effort in finding a correction to  my experiment that would bring the results into agreement with my own incorrect calculation for Au.  Thus I was well-equipped to reject this result outright, as did a number of others.

One possible solution is that the permittivity diverges as $\omega^{-2}$ as the frequency goes to zero.  This has led to the proposal of a generalized plasma model \cite{mosplas},
\begin{equation}\label{gp}
\epsilon_{gp}(i\xi)=\epsilon(i\xi)+{\omega_p^2\over \xi^2} ,
\end{equation}
where $i\xi$ represents the frequency along the imaginary axis, $\epsilon$ is the usual Drude model permittivity, for example, and $\omega_p$ is the so-called plasma frequency due to free electrons. Normally this expansion is assumed to be valid at very high frequencies, much above the resonances in the system of atoms and charges that comprise the plates.  However assuming the permittivity of this form brings back the contribution of the $TE$ $n=0$ mode, and apparently improves the agreement between theory and experiment.

There are consequences in a broader complex of phenomena when this generalized plasma model is introduced. In particular, if we consider the interaction of a low-frequency magnetic field with a material surface, by use of Maxwell's equation, it is straightforward to show that
\cite{jackson}
\begin{equation}\label{eddy}
-\nabla^2 \vec{H}={\omega^2\over c} \epsilon(\omega) \vec{H} ,
\end{equation}
which represents so-called eddy current effects, and can be easily extended to the complex frequency plane.  We see immediately that if $\epsilon$ diverges as $\omega^{-2}$ that at zero frequency,
\begin{equation}
-\nabla^2\vec{H}\propto \vec{H} ,
\end{equation}
which predicts that a static magnetic field will interact with an ordinary conductor in a manner different from universal diamagnetism.  Such an extra effect is not experimentally observed, as Eq. (\ref{eddy}) together with Eq. (\ref{cond}) is known to describe the non-diamagnetic interaction of low frequency fields with conductors.  So we are faced with discarding over a century of electrical engineering knowledge in order to explain a few 1\% level Casimir force experiments of questionable accuracy, and my own.  This is not acceptable.

The crux of the problem lies in the fact that at equilibrium, all electric fields at a surface of a conductor must terminate normal to the surface \cite{landl}. An electric field parallel to a surface implies a flowing current; such currents can exist in a transitory fashion as associated with a fluctuation as required for generating the Casimir force, but such fluctuations cannot occur with zero frequency.   For the $TE$ modes, the electric field is parallel to the surface, so at zero frequency $TE$ modes simply cannot be supported, assuming that equilibrium and zero frequency are equivalent.  We will return to this problem later in this review in relation to electrostatic calibrations.

This issue is, however, not yet settled as new precise experiments are required.  It is interesting that this effect becomes less pronounced at smaller separations, simply because the $n=0$ modes contribute a relatively smaller fraction to the total force.  For my own experiment \cite{skl97prl} the possibility  of a systematic error is becoming more and more apparent.  It should be emphasized, however, that AFM type experiments probe an order of magnitude smaller distance scale that the torsion pendulum experiments, and    the relative contributions of various effects are rapidly varying.

Work with AFMs and MEM type systems have demonstrated the difficulty of producing metal and other films, together with their characterization, that allows a comparison between experiment and theory at a level of better than 10\%.  For example, Svetovoy et al. \cite{optprops} show that the prediction of the Casimir force between metals with a precision better than 10\% must be based on the material optical response measured from visible to mid-infrared range, that the tabulated data is generally not good enough for precision work better than 10\% accuracy.  The issues of roughness are well-discussed in \cite{mohideenrev}, however, additional new work by de Zwol et al. \cite{dezwol} amplifies the problems of surface roughness particularly in determining the absolute separation.  It appears that the best prospect for determining the correct form of the permittivity function at zero frequency is to do a measurement at very large separations.  Indeed, problems of surface roughness correction virtually disappear for typical optical finishes at distances about 500 nm.  Above 2-3 microns, the difference between the force with and without the $TE$ $n=0$ mode approaches a factor of two.  Recent experimental work on Au films at Yale show that the Bostr\"om-Sernelius analysis is likely correct, but this work is at a very preliminary stage.

\subsection{The Detection of New Long Range Forces}\label{seclongrange}

In the mid-1980's, the question of the possible existence of a new so-called fifth force was suggested based on data from E\"otvos-type experiments \cite{fishbach}. Presently, interest in such forces is greater than ever due to possible modification of gravity as allowed by String Theory, and due to the observation of dark energy in the Universe which might be due to particles associated with new long range forces that could manifest themselves on many different length scales \cite{mostepdark}. The basic idea is that our four dimensional Universe is embedded in a space of more than 10 dimensions. Leakage of lines of force between the larger space and our four dimensional world could lead to a modification of the inverse square law, for example.  Although there is no specific prediction from a String theory, the possibility does exist in its context.

With the publication of my 1997 experimental result, I received many suggestions to analyze my experiment in light of an additional force that would appear along with the Casimir force, however I rejected these suggestions because my experiment was intended as a demonstration and any limit would be at the level of 100\% of the Casimir force.  Taken as a fraction of the gravitational field, my result was not particularly spectacular.  Nonetheless, other analyzed my experiment. Among the first to do so, in the context of a general review of limits on sub-centimeter forces, was Long et al. \cite{longprice} and earlier, with a more detailed analysis, was Klimchitskaya et al. \cite{klimfifth}.

The most ambitious recent work on this subject is by Decca et al. \cite{deccafifth} who  achieved an astounding accuracy without observing any anomalous effects.  Use of the proximity force theorem, to be discussed later in this review, to calculate the limits on a possible new force has been criticized.  The issue is that the proximity force theorem really only applies to a force that depends on the location of the body surfaces; the approximation is not valid for the volume integral required for calculating the anomalous force.  The applicability is addressed by Dalvit and Onofrio \cite{dalono} where corrections to the calculation in \cite{deccacalc} are pointed out.

Earlier work by Decca et al. \cite{deccaiso} appears as more reliable at constraining new forces. The technique developed here, a so-called isoelectronic method, relied on the properties of an Au film being independent of the substrate.  For different materials coated with Au films of identical optical characteristics and of sufficient thickness, the Casimir force should be the same.  In this work Au/Au and Au/Ge composites are compared, and the result is ``Casimir-less."  Techniques such as this appear as the most likely way to achieve the best sensitivity to new forces, however, unfortunately the minimum separation is limited by the Au film thickness, hence the later work \cite{deccafifth}.  It should be noted that use of a screening film to eliminate electrostatic forces and other background effect have been used in other ``fifth force" experiments for separations at the mm scale, but clearly the trick can be scaled down to distances limited only by the skill of the experimenter \cite{luther}.

\subsection{Modification of the Casimir Force}

The possibility of modification of the Casimir force is a topic of current great interest.  With the rising of nanotechnology, the need to control, modify, or make good use of the Casimir force is imperative as it is among the dominant forces affecting MEMs and NEMs.  At very short distances, at the atomic scale, the large-scale geometrical aspects of the surfaces become irrelevant, and the force becomes dominated by the van der Waals force between atoms comprising the plates; the atom-atom force along with roughness leads to stiction and friction.  At such short distances, the treatment of the plates in a continuum fashion fails. Any possibility to control either the short range or long range force can have enormous technological benefits.  These issues have generated renewed interest in measuring the Casimir force with improved precision, in applying it to nano-mechanical devices, and in controlling it.   In many instances, the attractive nature of the force leads to more problems that to solutions because, for example, it leads to irreversible sticking of the components in a nano-device.  There have been proposals to develop ``metamaterials" which provide a boundary condition that makes the force repulsive, but the extremely large frequency range of electromagnetic field modes that contribute to the force suggests that this is not possible \cite{dalvitmilonni}.

The internal sticking problem of MEMs, however, might be slightly overstated.  Recent commentary relating to this possible problem has been based on the work of Buks and Roukes \cite{bandr} where irreversible stiction was observed in MEMs devices.  In this work, the mechanical motion was monitored by use of an electron beam which caused the components of the MEMs to become highly charged.  Whether the irreversibility is really due to the Casimir force, or if it is due to charge surface interactions, remains an open question.  Nonetheless, it is agreed that a full understanding of the Casimir force, and its possible control, are central to the future of MEMs and NEMs engineering.

The prospects of engineering a coating that can significantly modify the Casimir force appear as dismal.  This is because the Casimir force is a ``broad-band" phenomenon.  Use of magnetic films has been suggested, but unfortunately ferromagnetic response does not extended into the near-infrared and visible spectrum that would be required to modify the Casimir force.

Recently, it has been demonstrated experimentally that a conductive oxide film, Indium-Tin Oxide (ITO) produces a Casimir force about half of that due to metals \cite{ito}. ITO has a number of interesting features, including transparency over the optical spectrum and chemical inertness.  Thus it appears as an interesting material from a nanoengineering viewpoint.

Casimir himself attempted to apply his namesake force to the electron, specifically to calculate the fine structure constant.  Casimir modelled the electron as a conducting ball of uniform charge that would contract due to the zero point energy of the external electromagnetic modes.  This force would be balanced by the space charge repulsion of the uniform charge density, when the conducting sphere of constant total charge was just the right diameter. The fine structure constant $\alpha\approx 1/137$ constant, which relates to the electron diameter, could then be determined from fundamental parameters along with a calculation of how the electromagnetic mode zero point energy changes as the sphere contracts \cite{milonni}.
However, Boyer subsequently found that the exterior spherical modes cause the sphere to expand \cite{boyer}. Boyer's result was interesting enough that it led to the exploration of the effects of geometry on the Casimir force.

The change in boundary conditions that had been considered cannot be realized experimentally; for example, if one cuts a conducting sphere in half and tries to measure the force between the hemispheres, the force is different from the stress outside the continuous conducting sphere-- simply because the two halves are now separated by a vacuum gap and there will be an attraction there, and because the structure of the surface modes is altered by the gap. Nonetheless, several experiments aimed at directly modifying the Casimir force have been performed in the last decade or so, and are continuing.

\subsection{Hydrogen Switchable Mirror}

An experiment with a surprising result employed a hydrogen switchable mirror, and a change in the Casimir force was sought when the mirror was switched between its low reflectivity and high reflectivity states \cite{capasso}.  The surprise was that no significant change in the Casimir force was observed with the switching, despite the rather dramatic change in the mirror from nearly transparent to highly reflecting.

The explanation of the null result likely lies in the construction of the mirror which has a very thin (5 nm) palladium layer to protect the underlying sensitive structure.  This layer tends to dominate the Casimir effect, even though the layer is about one-half of a skin depth for the frequencies that are affect by the hydrogen switching.  Other complications include the narrow spectral width of the mirror state which reduces the effect further, and the layered structure of the mirror--it is possible that the principal activity occurs in the deeper layers. In spite of these problems, hope remains that an effect on the Casimir force will be detectable \cite{demanh}.

\subsection{Geometrical Boundary Effects}

Until now, no significant or non-trivial corrections to the Casimir force due to boundary modifications have been observed experimentally.  As mentioned above, for the systems that had previously been considered such as the conducting sphere, it is not clear that an experimental measurement of the external stress is even possible. Cutting a sphere in half clearly changes the boundary value problem; it is unlikely that the two halves of such a sliced sphere will be repelled with a force that is given by the external stress on the sphere.

However, there are other possible ways to generate a geometrical influence on the Casimir force.  A conceptually straightforward way is to contour the surfaces of the plates at a length scale comparable to the mode wavelengths that contribute most to the net Casimir force.  For a plate separation $d$, the wavelengths that contribute most are $ \approx\pi d$.  This means that a surface nano-patterned at 400 nm length scale should show a significant geometrical effects for separations below 1 $\mu$m. Using such a system, Chan et al.   have produced a convincing measurement of  a non-trivial geometrical influence on the Casimir force \cite{chan}.

%%%%chan apparatus here
\begin{figure}[b]
\sidecaption
% For example, with the graphicx style use
\includegraphics[scale=.65]{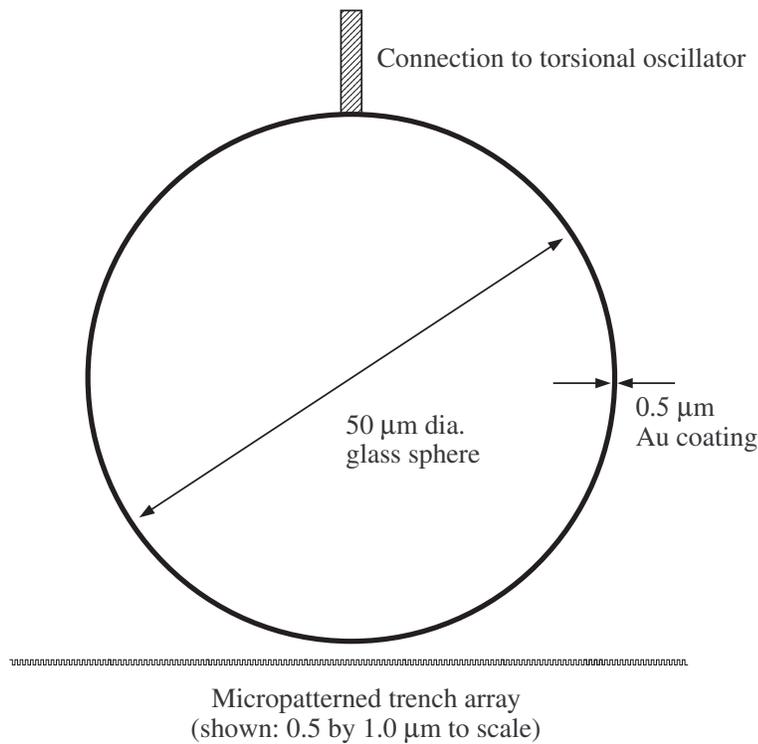}
%
% If no graphics program available, insert a blank space i.e. use
%\picplace{5cm}{2cm} % Give the correct figure height and width in cm
%
\caption{An approximately scaled schematic representation of the experiment of Chan et. al. The trench arrays, of varying width and depth, were made from the same doped p-type Si substrate.
(Public Domain, by S.K. Lamoreaux)}
\label{fig:1}       % Give a unique label
\end{figure}

These measurements, between a nanostructured silcon surface and a Au coated sphere, were made using a micromechanical torsional oscillator. The change in resonant frequency of the oscillator, as a function of separation between the Au sphere and the surface,  provided a measure of the gradient of the Casimir force.  The sphere, of radius 50 $\mu$m coated with 400 nm of gold,  was attached to one side of the oscillator that comprised a 3.5 $\mu$m thick, 500 $\mu$m square silicon plate suspended by two tiny torsion rods.  The sphere and oscillator was moved toward the nanostructured surface by use of a piezoelectric actuator.

Two different nanostructured plates, compared with a smooth plate, were measured in this work.  The geometry of the nanostructures, rectangular trenches etched in the surface of highly p-doped silicon, were chosen because the effects are expected to be large in such a geometry.  Emig and B\"uscher had previously calculated the effective modification of the Casimir force due to such a geometry, but for the case of perfect conductors \cite{emig}.  Even though the calculations were not for real materials, these theoretical results appeared as a reasonable starting point for a comparison with an experiment.

Although much progress has recently been made toward a realistic and believable accuracy and precision with which the Casimir force can be calculated for real materials \cite{optprops}, the problems associated with the well-known experimental variability of sputtered or evaporated films were avoided in the work of Chan et al. by comparing two different nanostructured plates with a smooth plate, all made from the same silicon substrate, and all using the same  Au coated sphere.  The trick is comparable to the Isoelectronic method described in Sec. \ref{seclongrange}. So even though ab initio calculations of the Casimir force for real material using tabulated optical properties cannot be accurate to better than 10\%, this problem was simply circumvented by the comparison technique.

The geometric modification of the Casimir force was detected by measuring a deviation from that expected by use of the Proximity Force Approximation (PFA), or the Pairwise Additive Approximation (PAA), both of which will be described later in this review.   The success of the PFA is so good that it suggests a means of detecting a geometrical effect.  Basically, the surface is divided into infinitesimal units, and it is assumed that the total Casimir force can be determined by adding the Casmir force, appropriately scaled by area, between surface unit pairs in opposite surfaces; this is the PAA. Thus, for the nanostructured surfaces, a 50\% reduction in force would be expected by the PAA, because the very deep trenches (depth $t= 2a\approx 1 \mu$m), etched as a regular array, were designed to remove half of the surface. As mentioned, two different trench spacings $\lambda$ were fabricated and measured, such that $\lambda/a=1.87$ (sample A) and $0.82$ (sample B), and compared to a smooth surface.  The Casimir force between the gold sphere and the smooth plate, as calculated from the tabulated properties of gold and silicon, taking into account the conductivity due to the doping, agree with the experimental results to about 10\% accuracy. For sample A, the force is 10\% larger than expected by the PAA, using the measured smooth surface force, and for sample B, it is 20\% larger, in the range $150 < z< 250$ nm. The deviation increases as $\lambda/
a$ decreases, as expected.

The theory of Emig and B\"uscher predicts deviations from the PAA twice as large as were observed. Nonetheless, the results of Chan et al. indicate a clear effect of geometry on the Casimir force. However,
much theoretical work remains to be done toward gaining a complete understanding of the experimental observations.  The already difficult calculations are made more so by the finite conductivity effects of the plates, and the sharp features of the trenches as opposed to the smooth simple sinusoidal corrugations. New calculational techniques have been developed that will allow reasonable accuracy calculations. Also a number of possible systematics associated with electrostatic effects were not fully investigated.

\subsection{Repulsive Casimir Effect}

The generalized Liftshitz formulation of the Casimir force allows for a material between the plates.  The force is thus altered from the case of a vacuum between the plates, and the effect can be calculated. It is easy to envision filling the space between the sphere and plate of a Casimir setup with a liquid and measuring the effects of replacing the vacuum.  A first experiment using alcohol between the plates was done by Munday et al. \cite{alcohol} where a substantial reduction in the force was observed compared to what is expected with vacuum between the plates.  The effects of Debye screening and other electrostatic effects were also thoroughly studied \cite{screening}.

Munday et al. extended their studies to a very interesting situation where the Casimir force becomes repulsive, by suitably choosing the permittivities of the plates and liquids. If the plates' material dielectric permittivities are  $\epsilon_1$ and  $\epsilon_2$, and the liquid between has  $\epsilon_3$, the force will be repulsive when $\epsilon_1 >  \epsilon_3 >  \epsilon_2$.  Of course, the permittivities are frequency dependent, so this relationship must hold over a sufficiently broad range of frequencies.

Perhaps a more familiar problem is the wetting of a material surface by a liquid.
In this case, one plate is replaced by air or vacuum so  $\epsilon_2=1$, and if the liquid permittivity is less than that of the remaining plate, the liquid spreads out in a thin film rather than forming droplets.  For example, liquid helium, which has a very small permittivity, readily forms a thin film because it is ``repelled" by the vacuum $( \epsilon_1 >  \epsilon_3 >  \epsilon_2 =1)$, and we say that the liquid wets the surface. On the other hand, liquid mercury which has a high effective permittivity does not wet glass $( \epsilon_1 <  \epsilon_3 >  \epsilon_2 =1)$.

Although there are many liquids that wet glass or fused silica, there are only a few sets of materials that will satisfy the requirement for a repulsive force between material plates. The set employed by Munday et al. was fused silica and gold, with bromobenzene as the liquid.  The experimental setup was based on an atomic force microscope (AFM) that was modified slightly for the detection of average surface forces rather than atomic-scale point forces.  For measuring the Casimir force, the sharp tip was replaced by a gold coated microsphere (diameter = 39.8 microns) which serves as the gold plate.  Using a spherical surface for one plate simplifies the system geometry, which is completely defined by the sphere radius and distance of closest approach from the flat fused silica plate.

A problem that all Casimir force experiments face is the system force calibration.  For this work and related work, a most clever calibration technique was devised. Because the fluid produces a hydrodynamic force when the sphere/plate separation is changed, and this force is linear with velocity, subtracting the force when the separation is changed at two different speeds produces the hydrodynamic force without any contribution from the Casimir force.  The hydrodynamic force thus measured, which can be calculated to high accuracy, provided the calibration.  In addition, this force, scaled to the appropriate velocity, was then subtracted from the force vs. distance measurement, yielding a clean measurement of the Casimir force.  The measurements spanned a range of 20 nm to several hundred nm, with the minimum distance limited by surface roughness, and the maximum distance limited by system sensitivity.  Various spurious effects were accounted for and shown to have no significant contribution within the statistical accuracy of the measurement.

Showing that it is indeed possible to produce and measure a repulsive Casimir force is important to both fundamental physics and to nanodevice engineering.  There has been much discussion of such forces as they will provide a means of quantum levitation of one material above another.  Even in a fluid, it will be possible to suppress mechanical stiction and make ultra-low friction sensors and devices.  It might be possible to ``tune" the liquid (e.g., by use of a mixture) so that at sufficiently large distances, the force becomes attractive, while being repulsive at short distances. This would allow objects to levitate above a liquid covered surface, for example.

\section{Approximations, Electrostatic Calibrations, and Background Effects}

Wittingly or unwittingly, many approximations have been included in all Casimir force experiments to date.  For example, most experiments employ the use an electrostatic force from accurately measured applied voltage for calibrations and the detection of spurious contact potentials between the plates.  The force is assumed to follow the form
\begin{equation}\label{cap}
F(d)={1\over 2}{\partial C(d)\over \partial d}V^2 ,
\end{equation}
where $C(d)$ is the capacitance between the Casimir plates, as a function of distance $d$ between them.  An exact calculation exists between a sphere and a plane, however, for most situations the so-called Proximity Force Approximation (PFA) can be used.  In the case of a plate with spherical surface with curvature $R$, with a distance $d$ at the point of closest approach to a plane surface, the force between the two plates is
\begin{equation}
F(d)=2\pi R {\cal E}(d) ,
\end{equation}
where ${\cal E}(d)$ is the energy per unit area between plane parallel surfaces that leads to the attractive force.

Briefly, the PFA was introduced by Deryagiun \cite{Deryagiun} to describe the Casimir force between a curved surfaces, and this approximation is known to be extremely accurate when the curvature is much less than the separation between the surfaces.   The PFA can be used beyond the Casimir force and has quite general applicability \cite{PFA}. The PFA is a special case of the Pairwise Additive Approximation (PAA) where the plate surfaces are divided into infinitesimal area elements, and the force is determined through a pairwise addition of corresponding elements.  The PFA and PAA work very well for electrostatic effects because, for a conductor (even poor) in equilibrium, the electric lines of force must be normal to the surface, otherwise currents would flow in contradiction to the assumption that the system is in equilibrium.

The use of a sphere and a flat plate vastly simplifies an experiment because the system is fully mechanically defined in terms of the point of closest approach and the radius of curvature of the sphere. For two flat plates the system is specified by two tilt angles, the areas, long-scale smoothness, and a separation, which all need to be defined, measured, and controlled. It is interesting to note that if the force is measured as a function of applied voltage in the sphere-plane configuration that the result should be
\begin{equation}
F(d)={\pi\epsilon_0 R\over d}V^2=\alpha V^2 ,
\end{equation}
where $\epsilon_0$ is the permittivity of free space, and $R$ is the radius of curvature of the spherical surface.  The absolute distance between the sphere and the plane surface is proportional to $\alpha^{-1}$ and this provides a means of determining the distance.

Even when the full form of the sphere-plane capacitance is used in Eq. (\ref{cap}), approximations still exist. Specifically, there are additional terms to the force given by Eq. (\ref{cap}) because the capacitance is in fact a tensor.  This can be easily seen, as when a charged sphere is bisected, the two halves repel each other, with a force
$$F={q^2\over 8R} , $$
where $q$ is the charge on the sphere \cite{landl} (Prob. 2, Sec. 5).  Note that this is the force for a fixed charge, which must be modified for a fixed voltage.  The point is that the two halves experience a force, even though their potential difference is zero; there are apparently additional terms that need to be added to Eq. (\ref{cap}).  As the geometry is not critical in this argument, we can conclude  that if the two plates of a Casimir experiment are at the same non-zero potential, there will be an additional force repulsive force between them. This sort of effect has not been considered at all.

The other problem that has received significant attention only recently is the effect of patch potentials on a conducting surface. The effect is well-known, and {\em is largest with clean samples} because when dirt is present, ions tend to accumulate at the boundaries between the patches, shielding the effect \cite{landl} (Sec. 23).

To date, every Casimir effect that has bothered measuring the contact potential as a function of distance has shown an apparent distance dependence of that potential. Various experiments are nicely reviewed in  \cite{staticsurvey}.  The basic essential problem manifests itself in
anomalous behavior in the electrostatic calibration of an experiment, for example, as experienced in \cite{kimanom}.  It was suggested that the anomalous effects that were observed are due to irregularities of the spherical surface.  Roughness effects \cite{rough} certainly can cause problems at short distances, but the possibility that the anomalous effects are due to simple geometrical effects is credibly discarded in \cite{staticsurvey}.

The contact potential is simply measured by finding a voltage potential difference $V_m$ between the two plates that minimizes the force given by Eq. (\ref{cap}).  $V_m$ is manifest as an asymmetry in the force between $\pm V$ applied between the plates.

For our measurements using Gemanium (Ge) plates \cite{ourge}, we were initially confused because a $1/d^{1.2}$ to $1/d^{1.5}$ force persisted when the electrostatic force was minimized at each distance.  Our initial conclusion was that there was a distance offset, as described in the next section, together with an uncompensated voltage offset. de Man et al. \cite{iannuzzi2009} have also observed a distance dependence of the contact potential, and concluded that it did not lead to any anomalies in their electrostatic calibrations, however, the measurements are at shorter distances than were used in the Ge experiment. In general, the relative electrostatic effect, compared to the Casimir force, should scale roughly as $(1/d)/(1/d^3)=d^2$. I will now tell the story of how we came to understand the results of our measurements using Ge plates.

%%%%photo here
\begin{figure}[b]
\sidecaption
% For example, with the graphicx style use
\includegraphics[scale=.65]{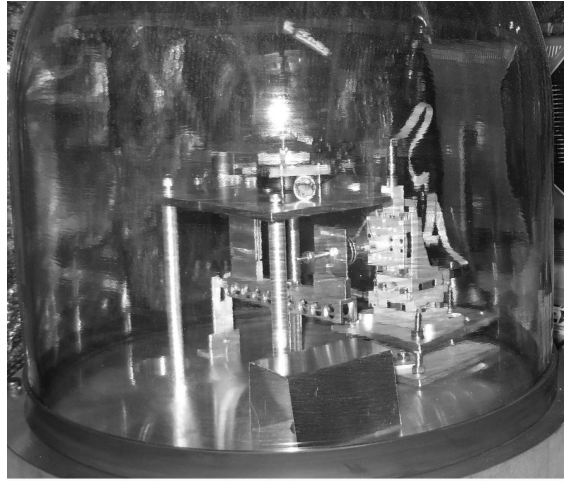}
%
% If no graphics program available, insert a blank space i.e. use
%\picplace{5cm}{2cm} % Give the correct figure height and width in cm
%
\caption{A photograph of the apparatus, in operation, used to measure the attractive force between Ge plates. The glass bell jar introduces some distortion; visible are the ``compensating plates" on the left of the torsion pendulum, and the plates (2.54 cm diameter) between which the Casimir force is measured, on the right. A ThorLab T25XYZ translation stage is used to position the ``fixed" plate.  The fine tungsten torsion wire is not visible. (Public Domain, by S.K. Lamoreaux)}
\label{fig:2}       % Give a unique label
\end{figure}

\subsection{Inclusion of the Debye Screening Length?}

In the early calibrations of our Ge plate Casimir experiment \cite{ourge}, we had a long-range background force that depended on distance not quite as $1/d$, as described above.  Our initial guess was that there was a distance offset in our calibrations due to penetration into the plates of the calibration electric field.
The problem is that a quasi-static
electric field can propagate a finite distance into a semiconductor (see, e.g., \cite{5});
this distance is determined by the combined consideration of
diffusion and field driven electric currents, leading to an
effective field penetration length (Debye-H\"uckel length)
\begin{equation}
\lambda=\sqrt{\epsilon \epsilon_0 kT\over e^2 c_t} ,
\end{equation}
where $c_t=c_h+c_e$ is the total carrier concentration, which for an
intrinsic semiconductor,  $c_e=c_h$. For intrinsic Ge
$\lambda\approx 0.6 \ \mu$m, while for a good conductor, it is less
than 1 nm.  $\lambda$ is independent of the applied field so long as
the applied field $E$ times $\lambda$ is less than the thermal
energy, $k_bT$ where $k_b$ is Boltzmann's constant.  In this limit,
and at sufficiently low frequencies and wavenumbers, thermal
diffusion dominates the field penetration into the material.  A
sufficiently low frequency for Ge would be $v_c/\lambda\sim 10$ GHz,
where $v_c$ is a typical thermal velocity of a carrier.

The potential in a plane semiconductor, if the potential is defined
on a surface $x=0$ is
\begin{equation}
V(x)=V(0)e^{-|x|/\lambda} ,
\end{equation}
where $\lambda$ is the
Debye-H\"uckel screening length, defined previously.

We are interested in finding the electrostatic energy between two thick Ge plates
separated by a distance $d$, with a voltages $+V/2$ and $-V/2$
applied to the back surfaces of the plates.  In this case, the field is
normal to the surface. After we find the energy per unit area, we
can use the proximity force approximation to get the attractive force
between a spherical and flat plate.

Let $x=0$ refer to the surface of the plate 1, and $x=d$ refer to
the surface of plate 2.  By symmetry, the potential at the center
position between the plates is zero.  The potential in plate 1 can
be written as
\begin{equation}
V_1(x)=V/2-(V/2-V_s) e^{-|x|/\lambda} ,
\end{equation}
and for the space between the plates
$$V_0(x)=-2 V_s x/d +V_s ,$$
where we assume the field is uniform.  $V_s$, the surface potential,
is to be determined.

We need only consider the boundary conditions in plate 1, which are
$$V_1(-\infty)=V/2 ,$$
$$V_0(0)=V_1(0) ,$$
(which has already been used)
$$\epsilon {d V_1(x)\over dx}\vert_{x=0}={d V_0(x)\over dx}\vert_{x=0}, $$
where the last two imply that $D=\epsilon E$ is continuous across
the boundary.

The solution is
\begin{equation}
V_s={V\over 2}\left({1\over 1+2\lambda/\epsilon d}\right).
\end{equation}
With this result, it is straightforward to calculate the total field
energy per unit area in both plates and in the space between the
plates. The result is
\begin{equation}
{\cal E}={1\over 2} {\epsilon_0 V^2\over d}\left[{y+y^2\over
(y+2)^2}\right] ,
\end{equation}
where the dimensionless length $y=\epsilon d/\lambda$ has been
introduced.  By expanding this result for small $y$, it can be easily seen that the effects appears as an apparent offset in the distance that is determined by measuring the capacitance between the plates.  For small voltages, this offset is approximately  $\lambda/\epsilon= 0.68/16\approx0.05\ \mu$m.

If $V-V_s$ is large compared to $k_bT$, the effective penetration
depth increases because the charge density is modified in the
vicinity of the surface.  The potential in the plates is no longer a
simple exponential, however one can define an effective shielding
length \cite{5}
\begin{equation}
{\lambda'\over \lambda}={|\Phi|\over \sqrt{e^{\Phi}+e^{-\Phi}-2}} ,
\end{equation}
where
\begin{equation}
\Phi={V-V_s\over k_bT} .
\end{equation}
Given that $k_bT=30$ meV, at
plate separations of order 1 $\mu$m for Ge this begins to be a large
correction when voltages larger than 60 mV are applied between the
plates, however, the potentials used in our experiment were far smaller.

We eventually realized that this effect is not present at very low frequencies; the lifetime of Ge surface states is on the order of milliseconds.  The lack of penetration of quasi-static fields into semiconductors was first observed in the development of the field effect transistor, and explained by Bardeen \cite{bardeen} as shielding due to surface states.  Again, in equilibrium, the electric field must enter normal to the plate surfaces, otherwise a current would be flowing in contradiction to the assumption of equilibrium.  Therefore, even on very poor conductors, charges rearrange to force any applied field to be perpendicular to the surface; when this situation is attained, the electric field terminates at the surface.  The boundary condition is that of a perfect conductor.

The presence of time-dependent surface states might be responsible for some of the anomalous electrostatic calibration effects observed by Kim et al. \cite{kimanom}. Particulary if there is a slight oxide coating on a metal surface, the surface states might not have enough time to reach equilibrium in the dynamic measurement system that was employed.  The relaxation times for trapped surface states can be many milliseconds.  However, the possibility that these sorts of states contribute to the anomalous effect is very speculative, and it is difficult to come up with an experiment to check this hypothesis.

As an aside, our consideration of this effect led us to the realization that the usual permittivity treatment of materials with non-degenerate conduction electrons is not correct, but must be solved in a different way than simply assigning a conductivity to the material \cite{carrierpaper}.  The discussion of this theoretical point is beyond the scope of this review.

\subsection{Variable Contact Potential}

It was recognized that a distance dependence of the minimizing potential would lead to extra electrostatic forces that are not necessarily zero at the minimizing potential \cite{myarxiv}. The force at the voltage which minimizes the force at each separation was thought to represent the pure ``Casimir" force between the plates. However, the applied voltage $V_a(d)$ required to minimize the (electrostatic) force is observed to depend on $d$, and is of the form (in the 1-50 $\mu$m range)
\begin{equation}
V_a(d)=a\log d +b ,
\end{equation}
where $a$ and $b$ are constants with magnitude of a few mV.  This variation leads to a long-range $1/d$-like potential for the minimized force.  An analysis suggests that this force is better described as $1/d^m$ where $m\approx 1.2-1.4$.

As we show here, the variation in $V_a(d)$ implies an additional force that increases as $1/d^{1.25}$, assuming that the voltage variation is due to the potential of the plates actually changing with distance.  Such changes could come about due to external fixed fields or potential variations associated with the plate translation mechanism, and is equivalent to having an adjustable battery in series with the plates.  We were unable to come up with a model that can give a sufficiently large effect based on interactions between, for example, the charge carriers. in the plates.  However, at sufficient sensitivity, it is likely that such effects will be important.

This analysis, while it predicts the correct form of the extra force, predicts that this force is negative or repulsive.
However, it is enlightening to go through the analysis, and this work will never be published elsewhere.
An understanding of the specific origin of the variation of applied minimizing potential $V_a(d)$ is not necessary to correct for the additional force that it causes, we simply need the experimentally determined $V_a(d)$, and assume it is tied to the plate positions.

We note further that $V_a(d)$ is not a measure of the contact potential, but the voltage which minimizes the force. We call the ``true" contact potential $V_c(d)$, which might depend on distance.

In performing our experiment, at each separation $d$, $V_a$ is varied and its value that minimizes the attractive force is determined.
\begin{equation}
{\cal E}(d)={1\over 2} C(d) (V_a+V_c(d))^2 ,
\end{equation}
where $C(d)$ is the capacitance between the plates, $V_a$ is the applied potential and is an independent variable, and $V_c(d)$ is the average weighted contact potential between the plates.

The force between the plates is given by the derivative of $\cal E$,
\begin{equation}
F(d)={\partial {\cal E}(d)\over \partial d}={1\over 2} {\partial C(d)\over\partial d}(V_a+V_c(d))^2+C(d)(V_a+V_c(d)){\partial V_c(d)\over \partial d} ,
\end{equation}

Now the minimum in the force is determined by the derivative with $V_a$:
\begin{equation}
{\partial F(d)\over \partial V_a}={\partial C(d)\over\partial d}(V_a+V_c(d))+C(d){\partial V_c(d)\over \partial d}=0 ,
\end{equation}
which determines $V_a(d)$, no longer an independent variable.  Thus,
\begin{equation}
{\partial V_c(d)\over \partial d}=-{1\over C(d)}{\partial C(d)\over\partial d}(V_a(d)+V_c(d)) ,
\end{equation}
which allows the determination of $V_c(d)$ when $V_a(d)$ is known.  The differential equation can be solved numerically, noting that at long distances $V_a(d)=-V_c(d)$, and that $V_c(d)$ become constant.

The electrostatic force between the plates at the minimized potential is given by
\begin{equation}
F(d)=-{1\over 2} {\partial \over \partial d} \left[C(d)(V_a(d)+V_c(d))^2\right].
\end{equation}
There are some nice features to this result.  First, if we apply a constant offset $V_0$ to $V_c(d)$, this effect is compensated by $V_a(d)-V_0$ which is easily seen as the relationship is linear.

Unfortunately, the sign of the effect indicates that it is repulsive, and thus is not the explanation of the long range force that persists at the minimizing potential as observed in our Ge experiment.

However, it should be emphasized that any precision measurement of the Casimir force requires verification that the contact potential is not changing as a function of distance, and if it is, a correction to the force as described here might very well exist.

\subsection{Patch Potential Effects}

It is often assumed that the surface of a conductor is an equipotential. While this would be true for a perfectly clean surface of a homogeneous conductor cut along one of its crystalline planes,  it is not the case for any real surface which can be polycrystalline, stressed, or chemically contaminated.
Experiments show that even with precautions for extreme cleanliness, typical surface potential variations are on the order of at least a few millivolts~\cite{LIGO}. This is most likely due to local variations in surface crystalline structure, giving rise to varying work functions and hence varying-potential patches. It is well known that the work function of a metal surface depends on the crystallographic plane along which it lies; as an example, for gold the work functions are 5.47~eV, 5.37~eV, and 5.31~eV for surfaces in the
$\langle 100 \rangle$, $\langle 110 \rangle$, and $\langle 111 \rangle$ directions respectively. This variation is most likely due to different effective electron masses, hence Fermi energies, for the different axes.

The means by which surface potential patches form is described in \cite{landl}, Sec. 22.  Briefly, When two conductors, A and B, of different work functions are brought into contact,  electrons flow until the chemical potential (i.e., the Fermi energy) in both conductors equalizes.  If we consider moving an electron in a closed path that moves from inside conductor A, across the boundary to inside conductor B, through the surface of B into the vacuum, back though surface A, and to the starting point, the total work must be zero in equilibrium.  If we take the contact potential difference between the conductors as $\phi_{ab}$, and the surface work functions as $W_a$ and $W_b$, for the total work to be zero we must have
$$\phi_{ab}=W_b-W_a ,$$
implying that the contact potential is simply the difference in the surface work functions.

It is straightforward to calculate the electric field energy of random patches, as has been done by Speake and Trenkel \cite{Speake2003}.
Consider two plane and parallel surfaces separated by a distance $d$.  Assume a potential $V=0$ at $x=0$,
while at $x=d$, $V=V_0\cos k y$.  It is easy to show that, in the region between the plates,
$$V(x,y)=V_0 \cos ky { e^{kx}-e^{-kx}\over e^{kd}-e^{-kd}}.$$

The field energy, per unit area is given by
$${\cal E}=\int_0^d \left[\left({\partial V\over \partial x}\right)^2+\left({\partial V\over \partial y}\right)^2\right] dx , $$
where we have used the fact that $\langle \cos^2ky\rangle=\langle \sin^2ky\rangle =1/2$ to do the $y$ integral.  Letting
$$u=e^{kx}-e^{-kx}\ \ \ \ dv=e^{kx}-e^{-kx} ,$$
so
$$du=k[e^{kx}+e^{-kx}]\ \ \ \ \ v={1\over k}[e^{kx}+e^{-kx}] ,$$
and integrating by parts
$$ \int_0^d[e^{kx}-e^{-kx}]^2dx={1\over k}[e^{2kx}-e^{-2kx}]\vert_0^d-\int_0^d [e^{kx}+e^{-kx}]^2dx.$$
The LHS is proportional to the field energy for $E_y$ while the last term on the RHS is proportional to (minus) the field energy for $E_y$.  We thus have
$${\cal E}=k{V_0^2\over 2} {e^{2kd}-e^{-2kd}\over [e^{kd}-e^{-kd}]^2}.$$

By use of the proximity force approximation, the (attractive) force between a flat surface and spherical surface is
$F(d)=2\pi R {\cal E}(d)$ where $R$ is the radius of curvature, where $d$ is the point of closest approach between the surfaces.  In the limit $kd\rightarrow 0$,
$$
F=2\pi R {V_0^2\over 4 d}\propto {1\over d}.
$$
This shows that when $kd \ll 1$ or $d \ll \lambda/2\pi$ where $\lambda$ is a characteristic length of a potential patch, the force goes as $1/d$.  This is what we expect from the PAA when the surfaces are very close.

There is an intermediate range where the force transforms from $1/d$ to exponential variation; at further distances, the force becomes a constant, as $\cal E$ does not vary with $d$. Between parallel plates, at long distances, the force is zero because the field energy does not change with separation. It is interesting to note this significant difference between the PFA result for a spherical surface and the result for parallel plates. As a constant force is in reality unobservable, this long distance force should be subtracted from the PFA result.

It should be noted that the field equations are linear, so we can add other $\cos(k'y),\ \cos(k'z)$ fluctuations, and the integral over $z,y$ leads to delta functions of $k-k'$. We can therefore rewrite the attractive force as an integral over $k_y, k_z$ where we have $V_{k_y}(y)+V_{k_z}(z)$ representing the amplitude spectrum in $k$ space of the surface fluctuations.  If we take $V_{k_y}(y)\sim V_{k_z}(z)$ and assume they are uncorrelated, the integral over $k_y, k_z$ leads to
$$F=\pi R V_{rms}^2 \int_0^\infty (2\pi k\ dk) (kS(k)) {e^{2kd}-e^{-2kd}\over [e^{kd}-e^{-kd}]^2} ,$$
where, by use of the Wiener-Khinchine theorem, $S(k)$ is the normalized cosine Fourier transform (in polar coordinates) of the surface potential spatial correlation function.

In order to compute the patch effect on the force in the sphere-plane configuration we make use of the
proximity force approximation. Just as in the case of roughness in Casimir physics \cite{rough},
one must distinguish between two PFAs: one is for the treatment of the curvature of the sphere (valid when
$d \ll R$, where $R$ is the radius of curvature), and the other one is the PFA applied to the surface patch distribution (valid when $k d \ll 1$). We assume that we are in the conditions for PFA for the curvature, but we keep $k d$ arbitrary. Then, the electrostatic force in the sphere-plane case
is $F_{sp}(d)= 2 \pi R {\cal E}(d)$, implying
\begin{equation}
F_{sp} =  2 \pi \epsilon_0 R \int_0^{\infty} dk \frac{k^2 e^{-k d}  }{\sinh(k d)} S(k).
\label{forcesp}
\end{equation}

There are a number of models that can be used to describe the surface fluctuations.  The simplest is to say that the potential autocorrelation function is, for a distance $r$ along a plate surface,
\begin{equation}
{\cal R}(r) = \left\{ \begin{array}{ll}
{V_{0}^2} & {\rm for}\ r \leq \lambda ,\\
0 & {\rm for}\ r > \lambda.
\end{array} \right.
\end{equation}
Then, by the Wiener-Khinchin theorem, the power spectral density $S(k)$ can be evaluated as the cosine two-dimensional Fourier transform of the autocorrelation function, which in our notation is \cite{stein}
\begin{equation}
S(k)={V_0^2}\lambda^2 {J_1(\lambda k)\over \lambda k},
\end{equation}
with $J_1$ the Bessel function of first kind.
The plane-sphere force is then given by, using  $k=u/\lambda$,
\begin{equation}
F_{sp}  = 2 \pi \epsilon_0 R \int_0^\infty du\ u {J_1(u) \over e^{2ud/\lambda}-1} .
\end{equation}
A numerical calculation shows that, for $d<.01\lambda$,
\begin{equation}
F_{sp}\approx {\pi\epsilon_0 R V_{0}^2\over d} ,
\end{equation}
suggesting that $V^2_{\rm rms}=V_0^2$, as expected.  For $50\lambda>d>\lambda$, the force falls with distance as $1/d^3$.

We see immediately that at short distances, there is a residual force due to patches that varies as $1/d$, and there is no minimizing potential that can compensate this effect.  It is, in a restricted sense, equivalent to having an oscillating potential between the plates; there is no way for a static field to compensate the oscillating field energy.

\begin{figure}[b]
\sidecaption
% For example, with the graphicx style use
\includegraphics[scale=.65]{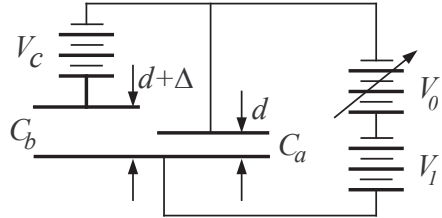}
%
% If no graphics program available, insert a blank space i.e. use
%\picplace{5cm}{2cm} % Give the correct figure height and width in cm
%
\caption{A toy model illustrating the mechanism for the generation of a distance-dependent minimizing electrostatic potential $V_m(d)$ and electrostatic residual force $F^{\rm el}_{\rm res}(d)$. (Public Domain, by S.K. Lamoreaux)}
\label{fig:3}       % Give a unique label
\end{figure}

As described in the last Section,
in our own work \cite{ourge} and in a number of other experiments \cite{kimanom,iannuzzi2009},
a distance-variation in the electrical potential minimizing the force between the plates has been observed.  It had been suggested already that this variation in contact potential can cause an additional electrostatic force, and an estimate was made for the possible size of the effect \cite{myarxiv}.  However, further experimental work shows that the model used in \cite{myarxiv}, where the varying contact potential is considered to be a varying voltage in series with the plates, does not reproduce our experimental results \cite{ourwork2009}.

A model that produces a residual electrostatic force consistent with our observations \cite{ourge} is shown in Fig. 3. In this figure, the two capacitors (short distance, $C_a(d)$, long distance $C_b(d+\Delta)$) create a net force on the lower continuous plate (setting $V_1=0$ initially),
\begin{equation}
F(d,V_0)=-\frac{1}{2} C_a' V_0^2 - \frac{1}{2} C_b'(V_0+V_c)^2,
\end{equation}
where
\begin{equation}
C_a'= {\partial C_a(d)\over \partial d};\ \ \  C_b'={\partial C_b(d+\Delta)\over \partial d} ,
\end{equation}
and $V_0$ can be varied, with $V_c$ a fixed property of the plates. The force is minimized when
\begin{eqnarray}
\left. {\partial F(d,V_0)\over \partial V_0} \right|_{V_0=V_m} = 0
& \Rightarrow & V_m(d)=-{C_b' V_c \over C_a'+C_b'} ,
\end{eqnarray}
implying a residual electrostatic force
\begin{eqnarray}
F^{\rm el}_{\rm res}(d) &=& F(d,V_0=V_m(d)) \nonumber \\
& =& -\left[C_a'+{C_a'^2\over C_b'}\right]{V_m^2(d)\over 2}
= -\left[ \frac{C_a' C_b'}{C_a'+C_b'} \right]{V_c^2\over 2} .
\end{eqnarray}
It is easy to take a case of parallel plate capacitors
($C_a'=-\epsilon_0 A/d^2$ and $C_b'=-\epsilon_0 A/(d+\Delta)^2$, where
$A$ is the area of each of the upper plates in Fig. 3, assumed to be equal;
hence, the lower continuous plate has area $2 A$)
and to show that there is a residual electrostatic force at the minimizing potential. Indeed, in such case,
\begin{eqnarray}
V_m(d) &=& - V_c \frac{d^2}{d^2 + (d+\Delta)^2} , \\
F^{\rm el}_{\rm res}(d) &=& \frac{\epsilon_0 A}{2} \; \frac{V_c^2}{d^2 + (d+\Delta)^2} .
\end{eqnarray}
Alternatively, in terms of $V_m(d)$ (up to $V_1$, see below), the force is
\begin{equation}
\label{fres}
F^{\rm el}_{\rm res}(d) = \frac{\epsilon_0 A}{2} \; \frac{V_m^2(d)[d^2 + (d+\Delta)^2]}{d^4} .
\end{equation}
Experimentally, $V_m(d)$ cannot be directly measured;  measurements can only
determine it up to an overall offset $V_1$ which arbitrarily depends on the sum of contact potentials in the complete circuit between the plates. Therefore the force should be written at proportional to $(V_m(d)+V_1)^2$ instead of simply $V_m^2(d)$, where $V_1$ is determined by a fit to experimental data, for example. In the limit $\Delta \gg d$, the residual force is proportional to $1/d^4$ in the plane-offset plane case here considered.

If we now consider the sphere-plane case, $C_a'(d)=-2\pi \epsilon_0 R/d$, and the denominator of Eq. (\ref{fres})
becomes $d^2$.  If we further consider the surface divided up into infinitesimal areas, each with a random potential, and integrate over the surface to get the net force, there is a further reduction of the power of $d$ in the denominator (just as in the proximity force approximation), leaving the sphere-plane force proportional to $1/d$.  This motivates writing the residual force as
\begin{equation}\label{reselec}
F_{res}(d)={\pi \epsilon_0 R\left[(V_m(d) + V_1)^2 + V_{rms}^2\right]\over d} ,
\end{equation}
where it is understood that $V_m(d)$ is experimentally measured, and $V_1$ is a fit parameter that represents a sort of surface average potential, plus circuit offsets (this equation is supported both by numerical studies and by our experimental results \cite{ourge,ourwork2009}, and is valid when $|V_1|>>|V_m(d)|$) as observed.  The last term in Eq. (\ref{reselec}) is the expected random (i.e., does not contribute to $V_m(d)$) patch potential force, but here should be thought as a fit parameter that reflects the magnitude of $V_{rms}$.  With this result, the long range force observed in our experiment could be explained, and our work with Ge was completed.  The agreement with theory is excellent, however, there is very little difference in theoretical prediction of the force with and without the $TE$ $n=0$ mode, so this work was not able to help with that controversy.

As a final note, the variations in surface potential could be a simple function of position on the conducting surface, for example, due to stresses or impurities within the samples. Alternatively, if there is a slight roughness to the surface, the peaks could have a different potentials that the valleys associated with surface irregularities.  This latter possibility appears to be a better model as we were unable to detect a variation in $V_m$ when the plates were moved relative to each other, which might be expected for positional surface patches.  However, the level of the surface fluctuations is quite small, and for example is beyond the range of state of the art Kelvin probes \cite{kelvinprobe}.  These issues need further investigation.

\section{Conclusions and Outlooks}

In many respects, we can consider the measurement of the Casimir force between surfaces as a mature field.  However, many open issues remain, particularly in the limits of accuracy that can be expected.  In recent years, we have seen a number of experiments claiming 1\% precision, but many counter claims that such accuracy is beyond what is possible due to finite knowledge of a plethora of corrections and required absolute calibrations.  Some open issues include the effects of finite conductivity on the contribution of the $TE$ $n=0$ surface mode; the usual Drude model of the permittivity of a metal suggests that this mode does not contribute at all to the force, reducing the force by a factor of two at large separations.  It is unclear whether additional short-range AFM type measurements will clear this problem up, as at short distances, the correction is relatively small.  Improved measurements at distances above a few microns would appear to offer the best prospects for bringing these issues to closure. Recent work with our torsion pendulum system at Yale seems to be in favor of the no-$TE$ $n=0$ mode, although the precision is not yet sufficient to make a strong claim. Over the next few months we hope to have new higher accuracy data analyzed.

The effects of patch potentials has not been fully investigated in all experiments to date. For example, in my 1997 experiment \cite{skl97prl}, an anomalous component to the $1/d$ force would result in an error in the distance determination, which only needed to be 0.1 micron to bring my experiment into agreement with the Bost\"om and Sernelius calculation.  Likewise, the boundary modification experiment of Chan et al. \cite{chan} did not consider in any obvious way excess forces due to electrostatic patch effects, which might be expected as substantial due to the sharp features of the etched silicon trenches, and will vary as $1/d^3$ in the limit of the separation much large than the trench spacing.  It is hard to imagine that such an effect is more than 10\% of the Casimir force, but some analysis and additional experiments are necessary to eliminate the possibility of such a systematic effect.

In any case, a reasonable ultimate experimental goal is the attainment of 1\% agreement between theory and experiment, in terms of true accuracy; this is not a question of simple precision.  Hopefully the readers of this review will realize the complexity and difficulty of the challenge presented by this goal.

\section{Acknowledgements}

I thank my colleagues and collaborators W.-J. Kim, A.O. Sushkov, H.X. Tang, and D.A.R. Dalvit for many fruitful discussions that led to the understanding of our Ge experiment, and to deeper understanding of the Casimir force in general.  I also thank R. Onofrio and S. de Man to a number of discussions over the last few years that were helpful in clarifying a number of issues.  SKL was supported by the DARPA/MTO Casimir Effect Enhancement project under SPAWAR contract number N66001-09-1-2071.

\end{document}